\documentclass[a4paper,11pt]{article}
\usepackage{pos}

\newcommand{\la}{\lambda}

\title{Predictions for neutron star mergers from the gauge/gravity duality}

\author*[a,b]{Matti J\"arvinen} 

\affiliation[a]{Asia Pacific Center for Theoretical Physics,   Pohang 37673,   Korea}

\affiliation[b]{Department of Physics, Pohang University of Science and Technology,   Pohang 37673,   Korea}

\emailAdd{matti.jarvinen@apctp.org}

\abstract{The gauge/gravity duality, combined with information from lattice QCD, nuclear theory, and perturbative QCD, can be used to constrain the equation of state of hot and dense QCD. I discuss an approach based on the holographic V-QCD model, which includes both nuclear and quark matter phases, separated by a first order phase transition. By using this model in state-of-the-art simulations of neutron star binaries, I study the formation of quark matter during the merger process, and its effect on the threshold mass for prompt collapse into a black hole.}

\FullConference{The XVIth Quark Confinement and the Hadron Spectrum Conference (QCHSC24)\\
 19-24 August, 2024\\
 Cairns Convention Centre, Cairns, Queensland, Australia\\}

\begin{document}
\begin{flushright}
APCTP Pre2025 - 007 
\end{flushright}

\maketitle

\section{Introduction}

\begin{figure}[ht!]
\centering
\includegraphics[width=0.4\textwidth]{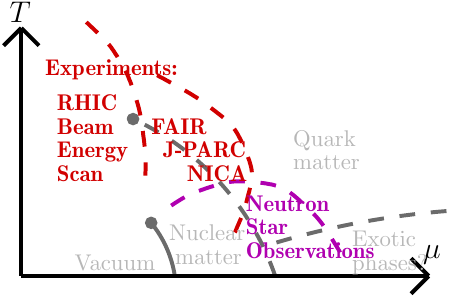}\hspace{12mm}%
 \includegraphics[width=0.4\textwidth]{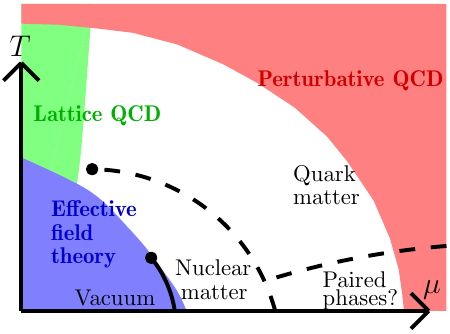} 
\caption{\label{fig:pd_sketch}The phase diagram of QCD as a function of the quark (or baryon) number chemical potential $\mu$ and the temperature $T$. Left: sketch of the ranges of data covered by ongoing and future heavy-ion experiments and neutron star observations, compared to a sketch of the relevant phases. Right: A sketch of the extent of validity for the main first-principles theoretical methods. Figure from~\cite{Jarvinen:2021jbd}.
}
\end{figure}

Experimental efforts are expected to improve our knowledge of 
the phase diagram of QCD drastically in the coming years. There is already plenty of data from the Beam Energy Scan program at the Relativistic Heavy-Ion Collider (RHIC) in Brookhaven, which focuses on the region near the deconfinement-confinement crossover in QCD and locating the critical point in the $(\mu,T)$ phase diagram. Future heavy-ion experiments, such as the experiments planned at the Facility for Antiproton and Ion Research (FAIR) and at Japan Proton Accelerator Research Complex (J-PARC), will extend the probed region of the phase diagram towards higher densities, see Fig.~\ref{fig:pd_sketch} (left).

These experiments will be complemented by observations of neutron stars. Apart from measurements of isolated pulsars, multi-messenger signals neutron star mergers will provide information about QCD at low temperatures and at densities reached at neutron star cores, i.e., densities comparable to the nuclear saturation density. Together with heavy-ion collisions, such observations are expected to dramatically increase our knowledge of QCD phase diagram, in particular in the region of the deconfinement and nuclear to quark matter transitions.

Given that lots of new data is expected in near future, it is also timely to try to improve theoretical predictions in the relevant region of the phase diagram. Namely, available results from first-principles calculations do not cover this region (see the right panel in Fig.~\ref{fig:pd_sketch}): Lattice analysis is limited to low densities (i.e., $\mu/T \lesssim 1$, if $\mu$ is the quark number chemical potential) due to the sign problem. Effective methods based on describing QCD by using hadronic degrees of freedom, such as chiral perturbation theory, are only reliable at low temperatures and densities (i.e., temperatures below the crossover temperature and densities around and below the nuclear saturation density). At high temperatures and densities, perturbative QCD works, but the errors are only controllable at much higher temperatures or densities than those found (for example) in neutron star mergers. This leaves a gap in the first principles analysis at intermediate densities, white region of Fig.~\ref{fig:pd_sketch} (right), which includes the expected nuclear to quark matter transition and the regions most important for physics of neutron stars, core collapse supernovae, and neutron star mergers. 

In the absence of first-principles results, the region of intermediate densities has been studied by using modeling, including nuclear matter models with various nucleon-nucleon potentials, mean-field theory, and Nambu-Jona-Lasinio models. A modern approach which may nicely complement these studies (see~\cite{Ghoroku:2013gja,Hoyos:2016zke,Mamani:2020pks,Kovensky:2021kzl}) is the gauge/gravity duality, which maps strongly coupled problems in four dimensional field theory to tractable computations in five dimensional classical gravity. 

When applied to QCD, the gauge/gravity approach follows some of the general principles of the original AdS/CFT correspondence for the $\mathcal{N}=4$ super-Yang-Mills. For example, the geometric interpretation is that the field theory lives at the boundary of a higher dimensional space. 
However, as QCD is not conformal, but confining, there will be some obvious differences: The five-dimensional dual geometry is not the anti-de-Sitter space, but some more general geometry.  The renormalization group flow is implemented by the evolution of the geometry as a function of the holographic coordinate. 
The dictionary is defined by mapping the operators of QCD to classical fields on the gravity side, and implemented explicitly by stating the equivalence between the classical gravity partition function and the generating functional of QCD. 
Different phases map to qualitatively different geometries.   
Thermal equilibrium states in QCD are implemented by planar black hole geometries, with the horizon extending to all space-time coordinates. The thermodynamics of field theory maps to the black hole thermodynamics of this geometry.

Since there are many other approaches, it is important to motivate introducing the gauge/gravity duality as an additional model at finite temperature and density. 
There are several arguments which support using this method. First, gauge/gravity is by definition a strongly coupled method. The failure of first-principles approaches in the white region of Fig.~\ref{fig:pd_sketch} (right) is due to the dynamics being strongly coupled. Therefore, it is natural to expect that holography could work better than other approaches in this region. Second, in gauge/gravity duality, it is usually easy to describe different phases in the same footing or even within a single model. In the optimal situation, they are given by different geometries for the same gravity action. This means that it may be easy to create an overarching holographic model, that describes the whole phase diagram, or most of it, and gives precise predictions also for the phase transition lines and the strengths of the various transitions. Third, as it turns out, the predictions from gauge/gravity duality are in good agreement with know constraints. Examples will be given below. This is quite nontrivial, as a precise gravity dual for QCD is not known and it is not even known whether such a dual exists in some well-defined sense.

\section{Holographic equation of state}

In this section, I focus on results presented in~\cite{Demircik:2021zll}. The main goal of this article was to construct an equation of state (EOS) based on strongly coupled gauge/gravity approach to be used in state-of-the-art neutron star mergers. However the resulting general-purpose EOS can naturally also be used to describe isolated neutron stars as well as an input for simulations of core-collapse supernova explosions. Moreover, the construction also covers the region relevant for heavy-ion collisions.

The construction follows a hybrid approach, where the use of holography is combined with other methods in regions where holography is not useful. This allows us to build a model, which pretty much agrees with all available data and constraints  from theory, experiments, and observations. As for the holographic model, we choose to use the V-QCD model~\cite{Jarvinen:2011qe,Jarvinen:2021jbd}. This is the only available model on the market which is versatile enough to work in all regions where we apply holography. However some parts of the construction could also be done by using the so-called Einstein-Maxwell-Dilaton ~\cite{DeWolfe:2010he,Knaute:2017opk,Critelli:2017oub,Cai:2022omk} or Witten-Sakai-Sugimoto~\cite{Sakai:2004cn} models.

The main ingredients of the model  (to be discussed in more detail below) are:
\begin{enumerate}
 \item {\em Holographic quark matter.} We use the V-QCD model to describe the phase diagram at high $\mu$ and $T$, i.e., the quark-gluon plasma and the (unpaired) quark matter phases.
 \item {\em Holographic dense nuclear matter.} We use a simple approach within the V-QCD model to describe cold and dense nuclear matter. However the temperature dependence is estimated by using a different van der Waals approach in order to achieve a state-of-the-art model.  
 \item {\em Low density QCD matter.} We use ``traditional'' approach, in this region covering low density nuclear matter and low-temperature meson gas phases, where obtaining feasible predictions from holography is challenging.
\end{enumerate}

\subsection{Holographic quark matter}

The quark matter phase is described by the V-QCD model~\cite{Jarvinen:2011qe}. It is a fusion of the improved holographic QCD model~\cite{Gursoy:2007cb,Gursoy:2007er}, which controls the gluon degrees of freedom, and a mechanism to describe quarks in terms of a tachyonic Dirac-Born-Infeld action~\cite{Bigazzi:2005md,Casero:2007ae}. The backreaction of the quarks to the glue is fully included in the Veneziano limit, where both the numbers of colors $N_c$ and the number of flavors $N_f$ are taken to infinity keeping their ratio fixed. The ``V'' in the name of the model refers to this limit. Note that the model is firmly rooted in string theory: the gluon sector arises from five-dimensional noncritical string theory and the quark sector arises from a space-filling pair of D4 branes. However in the end, in order to precisely describe QCD, one resorts to a bottom-up approach where details are determined by comparing to QCD data directly.

Next I write down the model action for chirally symmetric quark matter, i.e., at zero quark mass and in the absence of chiral symmetry breaking. Apart from gravity, the model contains a scalar $\lambda=e^{\phi}$, where $\phi$ is the dilaton field. This field is dual to the $G_{\mu\nu}^a G^{\mu\nu}_a$ operator in QCD, where $G$ is the gluon field. 
The dictionary also includes a gauge field $A_\mu$ dual to the vectorial current $\bar \psi \gamma_\mu \psi$.

The action is
\begin{align}
\begin{aligned} \label{eq:SVQCD}
  \mathcal{S}_\mathsf{V-QCD} &= {N_c^2} M^3  \int d^5x\, \sqrt{g}\left[R-\frac{4}{3}\frac{(\partial \la)^2}{\la^2} +{V_g}(\la)\right] \\
 & \ \ - {N_fN_c} M^3 \int d^5 x\ {V_{f0}}(\la) 
 \sqrt{-\det(g_{\mu\nu}+ 
 {w}(\la) F_{\mu\nu})} &
\end{aligned}
\end{align}
where $M$ is the five-dimensional Planck mass, and $F_{\mu\nu}$ is the field strength tensor for $A_\mu$. Here the first line describes the improved holographic QCD model (the gluon sector) and the second line is the DBI action (the quark sector), which is fully backreacted to the gluon sector. Importantly, the action contains three potentials $V_g$, $V_{f0}$, and $w$, which will be determined by comparing to QCD data (see the review~\cite{Jarvinen:2021jbd}). 
The most important matching with QCD is carried out at intermediate values of $\la$, where the potentials are determined by comparing to lattice data for the thermodynamics of QCD (with 2+1 flavors) or the particle spectrum~\cite{Gursoy:2009jd,Jokela:2018ers}. Various fits of the V-QCD model exists in the literature, depending on which data is stressed in the fit~\cite{Jokela:2018ers,Amorim:2021gat,Jarvinen:2022gcc}. In this talk I present results based on the fit done in~\cite{Jokela:2018ers,Ishii:2019gta}, as it leads to most accurate description of the QCD thermodynamics.

The comparison of the model to lattice data (and the computation of the EOS in regions where lattice data is not available) is done by 
constructing numerically charged planar black hole solutions in the gravity theory~\eqref{eq:SVQCD}, and extract the results from the thermodynamics of the black holes.

\subsection{(Holographic) nuclear matter}

It is well known how nucleons can be added in holographic models:  
the construction boils down to specific topological soliton solutions of gauge fields~\cite{Kim:2006gp,Hata:2007mb}. These solutions are often called instantons, because they are similar to the instantons appearing in Yang-Mills, expect that they are localized in the holographic coordinate instead of time. 

Such solitons have been recently constructed in the V-QCD model~\cite{Jarvinen:2022mys,Jarvinen:2022gcc}. 
However, constructing a nuclear matter configuration out of these solitons is hard: Already solving the model for a single baryon is a nontrivial numerical task. 
Considering multi-instanton solutions with instanton interactions is extremely challenging. In this talk I will therefore consider a simpler approach, which can be understood to correspond to instantons smeared over spatial directions. This leads to a homogeneous gauge field configuration, which may be a good approximation of the exact inhomogeneous solution at high densities~\cite{Rozali:2007rx,Li:2015uea,Ishii:2019gta,Kovensky:2021ddl,Bartolini:2022rkl}. 
 
In order to write down the ``homogeneous Ansatz'' one needs to set $N_f=2$ and consider a more general gauge fields than above, i.e., fields $A_{\mu}^{ij}$ transforming under the non-Abelian  U$(2)_V$. 
The Ansatz is then written as  $A_{k} (r) = h(r) \sigma_{k}$, where $r$ is the holographic coordinate and $\sigma_{k}$ are Pauli matrices. Note that the spatial Lorentz index in $A_k$ is linked to the adjoint flavor index. 

However, it turns out~\cite{Ishii:2019gta} that the resulting nuclear matter EOS has an obvious weakness typical for confined phases in holography: it is independent of temperature. 
This is perhaps a reasonable zeroth-order approximation, but clearly not sufficient to construct a state-of-the-art EOS. We fix this by introducing a van der Waals model, i.e., gas of protons, neutron and electrons with an excluded volume correction and an interaction potential. The potential is chosen such that the model precisely agrees with the V-QCD nuclear matter prediction at zero temperature, so that the van der Waals model gives a natural extrapolation of the V-QCD EOS to nonzero temperature~\cite{Demircik:2021zll}. 

In the low density regime for nuclear matter using holography is tricky, because one cannot avoid dealing with instantons.  
Because of this we use a well-established nuclear theory EOS at low density. We choose the DD2 version~\cite{Typel:2009sy} of the Hempel-Schaffner-Bielich equations of state,  HS(DD2)~\cite{Hempel:2009mc}. 
In addition, we include the pressure of free mesons, using the masses from the particle data group, in order for the low-temperature hadronic EOS to smoothly match with the V-QCD quark-gluon plasma EOS at higher temperatures.

So far we have only discussed the dependence of the model on temperature and density, and essentially restricted to isospin-symmetric description both holographic nuclear and quark matter. For state-of-the-art neutron star EOS, one also needs the dependence on the electron fraction,\footnote{Note that the van-der-Waals construction and the HS(DD2) also include electrons, and we also add the free electron pressure in the model for quark matter.} which allows the description of matter at and out of $\beta$-equilibrium. In order to achieve this we used for the electron fraction dependence a simple approximation in the holographic computation in the quark matter phase and the HS(DD2) model in the nuclear matter phase. 

\begin{figure}[ht!]
\centering
\includegraphics[width=0.4\textwidth]{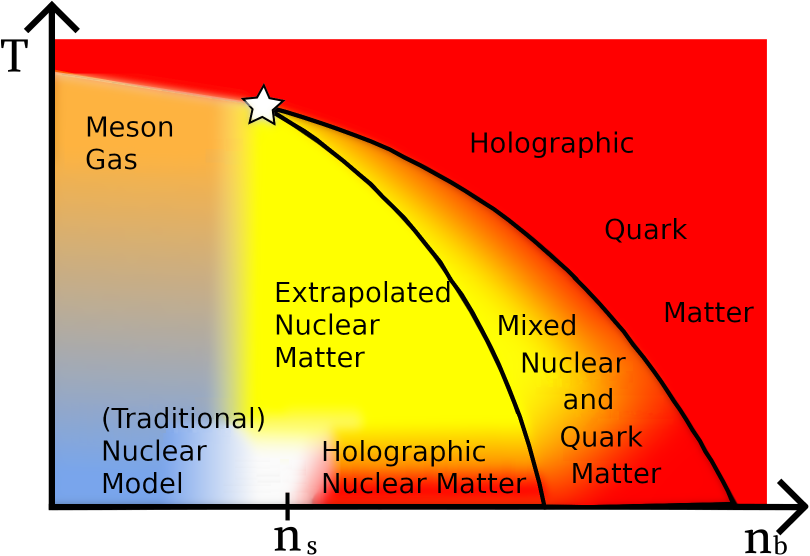}\hspace{12mm}%
\includegraphics[width=0.45\textwidth]{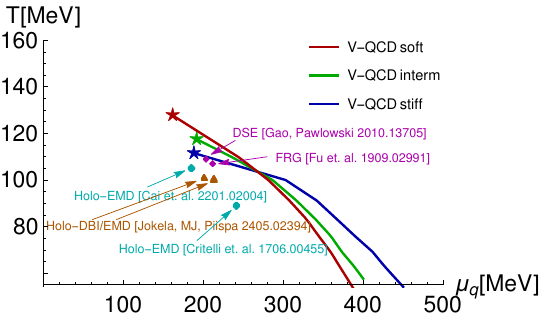}
\caption{\label{fig:pd_hybrid}. Left: Building blocks of the hybrid model and the phase diagram as a function of the baryon number density and temperature. Figure adapted from~\cite{Demircik:2021zll}. Right:
The critical point in our V-QCD approach (stars) compared to other results for the critical point.}
\end{figure}

\subsection{Combining the building blocks}

Combining the building blocks of the model, the resulting phase diagram is sketched in Fig.~\ref{fig:pd_hybrid} (left). Note that the model covers region relevant both for neutron star mergers and heavy-ion collisions. While the idea of gluing together EOSs from different approaches has been used frequently in the literature, our model is one of the most extensive attempts available. It is in agreement with basically all known experimental, observational and theoretical constraints. While the main motivation of our construction was to describe the region of dense matter, we also obtain predictions for the location of the critical point, as sketched in Fig.~\ref{fig:pd_hybrid}. Three variants of the resulting EOS (soft, intermediate, and stiff), given by different fits to lattice data, have been published in the CompOSE database~\cite{Typel:2022lcx} of EOSs. 

The success of this construction arises from several details that work remarkably well:
\begin{enumerate}
 \item A precise fit to data lattice to thermodynamics of the Yang-Mills theory and full QCD (with 2+1 flavors) was possible~\cite{Jokela:2018ers,Ishii:2019gta} with a smooth, monotonic Ansatz for all potentials in the action~\eqref{eq:SVQCD}. This is remarkable in part because after requiring agreement qualitative features of QCD (such as confinement) there is relatively little room in adjusting the potentials.
 \item The predictions for the extrapolated quark matter EOS at high density and small temperature, i.e., the region far beyond the convergence radius of any expansion of the small density lattice data, is feasible despite tight theoretical and observational constraints~\cite{Jokela:2018ers,Jokela:2020piw,Jokela:2021vwy,Demircik:2021zll}.
 \item The EOS for dense nuclear matter turns out to be stiff: the speed of sound clearly exceeds the value $1/\sqrt{3}$ of conformal theories. This is phenomenologically important as it makes it easy to pass observational constraints from measurements of neutron star masses.
 \item As both nuclear and quark matter are described by the same holographic theory, we can obtain sound predictions for the location and the strength of the nuclear to quark matter phase transition. The transition is strong, with latent heats $\sim 1000$ MeV fm$^{-3}$ at low temperature. This excludes the possibility of having stable quark matter cores in isolated neutron stars~\cite{Jokela:2018ers}.
\end{enumerate}

Another observation is that the predictions from the holographic model seem to agree remarkably well with those obtained by using the functional renormalization group (FRG) approach. In particular, our predictions for the location of the critical point~\cite{Demircik:2021zll} are close to those obtained by the FRG~\cite{Fu:2019hdw} and Dyson-Schwinger~\cite{Gao:2020fbl} computations, see Fig.~\ref{fig:pd_hybrid} (right). They are also close to the critical point obtained in simpler holographic constructions, i.e., by Einstein-Maxwell-dilaton~\cite{Critelli:2017oub,Cai:2022omk,Jokela:2024xgz} and Dirac-Born-Infeld~\cite{Jokela:2024xgz} actions. The slightly higher critical temperature than in other holographic models may stem from the inclusion of the nuclear matter component.

The agreement with the FRG is not restricted to the location of the critical point, but our EOS also seems to agree with computations in this approach. The speed of sound at zero temperature~\cite{Ishii:2019gta,Jokela:2020piw} is close to the result from the FRG both in the nuclear matter~\cite{Drews:2016wpi} and quark matter~\cite{Otto:2019zjy} phases.

\section{Holographic neutron star mergers}

Binary neutron star mergers are significant sources of both gravitational and electromagnetic radiation. The multi-messenger signals from the GW170817 event~\cite{TheLIGOScientific:2017qsa} are already giving us nontrivial information about the equation of state of dense QCD matter~\cite{Annala:2017llu}.

As a function of the total mass of the binary, three qualitatively different scenarios for the stages of the merger event are possible (see, e.g.,~\cite{Baiotti:2016qnr}). At high masses (as compared to the solar mass), the neutron stars promptly collapse into a black hole (leaving behind a torus of matter) right after they merge. At lower masses, a hypermassive neutron star is formed, followed by a collapse into a black hole in the millisecond or second timescale. At even lower masses, no black hole is formed at the merger, but a supramassive neutron star is formed. However, it may later collapse in to a black hole due to cooling of the star and slowing down of the rotation. 

We simulate neutron star mergers by numerically solving Einstein equations and relativistic hydrodynamics. As a physics input in the simulations, we choose the hybrid V-QCD EOS discussed above. Moreover, we use the Frankfurt University/Kadath code for the initial configuration~\cite{Papenfort:2021hod}, and Frankfurt/Illinois code for the time evolution~\cite{Most:2019kfe}, using the Einstein toolkit framework.
 
\subsection{Production of quark matter}

As the first example, we analyze quark matter production in neutron star mergers~\cite{Tootle:2022pvd}. This is a natural application of our EOS model, because it describes nuclear and quark matter in the same footing and therefore gives predictions for the phase transition between these phases.

We ran simulations with the total mass of the system matching with the GW170817 event, varying the mass ratio of the constituents and the variant (soft, intermediate, or stiff) of the V-QCD hybrid EOS. The simulations with the soft EOS show a hypermassive neutron star phase with a collapse to a black hole about 10 milliseconds after the merger. The simulations with the intermediate and stiff EOSs do not show a collapse within the time range of the simulation. 
Using the simulation results, we identified and analyzed three different stages in the quark matter production that we coin hot quarks, warm quarks, and cold quarks (see also~\cite{Prakash:2021wpz}).

The hot quark stage appears right after the merger: the matter heated in the collision is driven above the phase transition line of Fig.~\ref{fig:pd_hybrid}, so that the quark matter is formed in the hottest regions of the hypermassive star. Warm quarks are produced in the next stage. In this stage, complicated oscillating dynamics drives formation of quark matter in regions of the star which are not the hottest nor the densest. Finally, in the third stage cold quarks are formed. That is, a dense and cold quark core forms if the density of the center of the star exceeds the critical density of the phase transition at zero temperature. For our EOSs, the formation of the cold quark core always triggers an immediate collapse into a black hole, as the soft quark matter EOS cannot support a stable quark core.

The collapse to the black hole can be read off from the gravitational wave signal, so that the quark matter formation can in principle be observed this way. Indeed, our simulations for the EOS where the quark matter phase has been removed by hand show significantly delayed collapse compared to the simulations with quark matter. However, it is difficult to tell from the gravitational wave signal directly what the cause of the collapse was. We also remark that our simulations with the soft EOS showed a rapid collapse to a black hole (within ten milliseconds) whereas the electromagnetic signal from GW170817 has been argued to imply a much slower collapse, about one second after the merger~\cite{Gill:2019bvq}. This observation disfavors the soft EOS model.

\subsection{Prompt collapse to a black hole}

In a recent article~\cite{Ecker:2024kzs} we carried out a systematic analysis of the threshold mass of the binary for the prompt black hole collapse by using the hybrid V-QCD EOSs. For this analysis we ran simulations of equal-mass binaries over a range of masses using the three V-QCD variants.

\begin{figure}[ht!]
\centering
\includegraphics[width=0.37\textwidth]{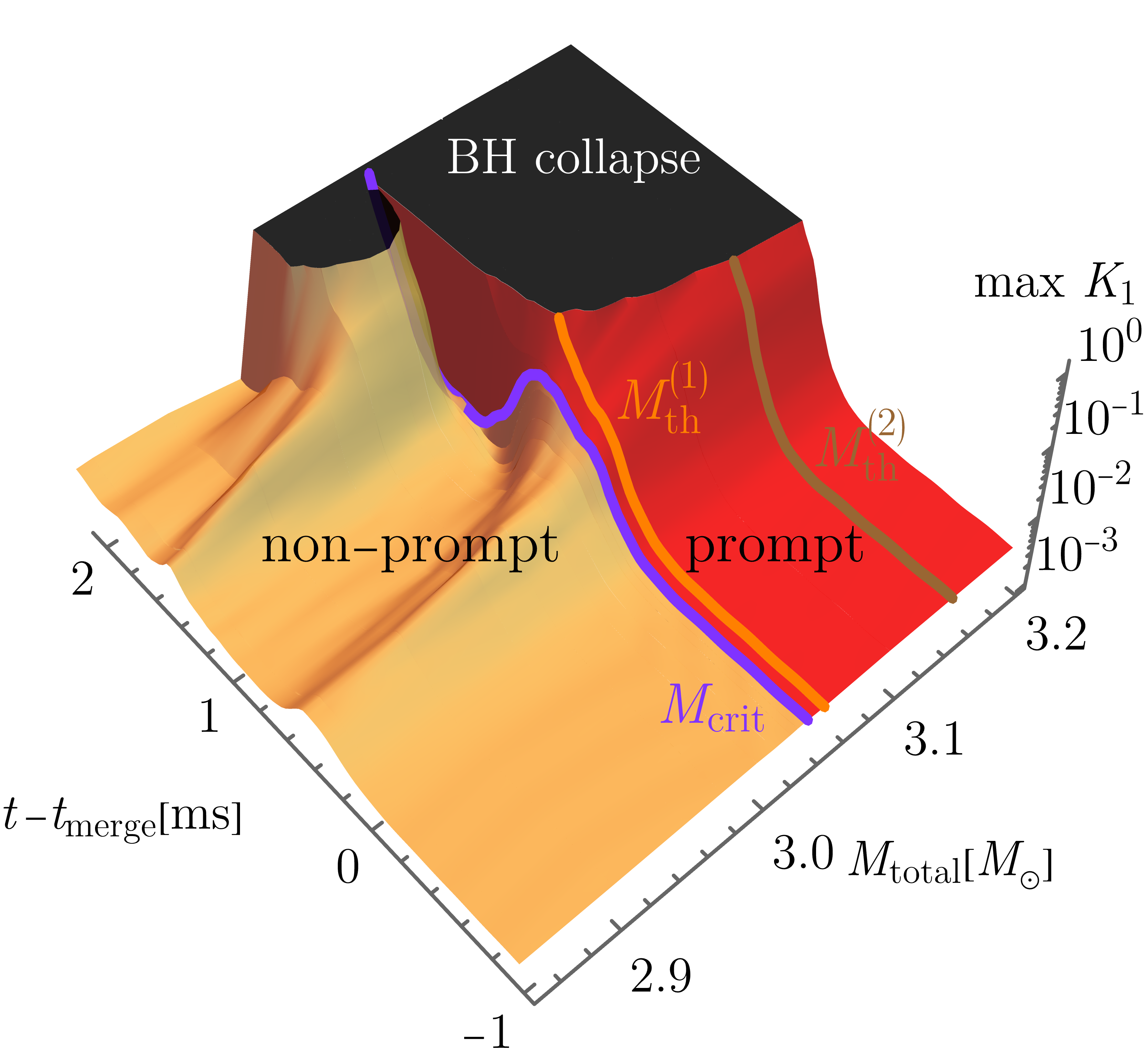}\hspace{16mm}
\includegraphics[width=0.37\textwidth]{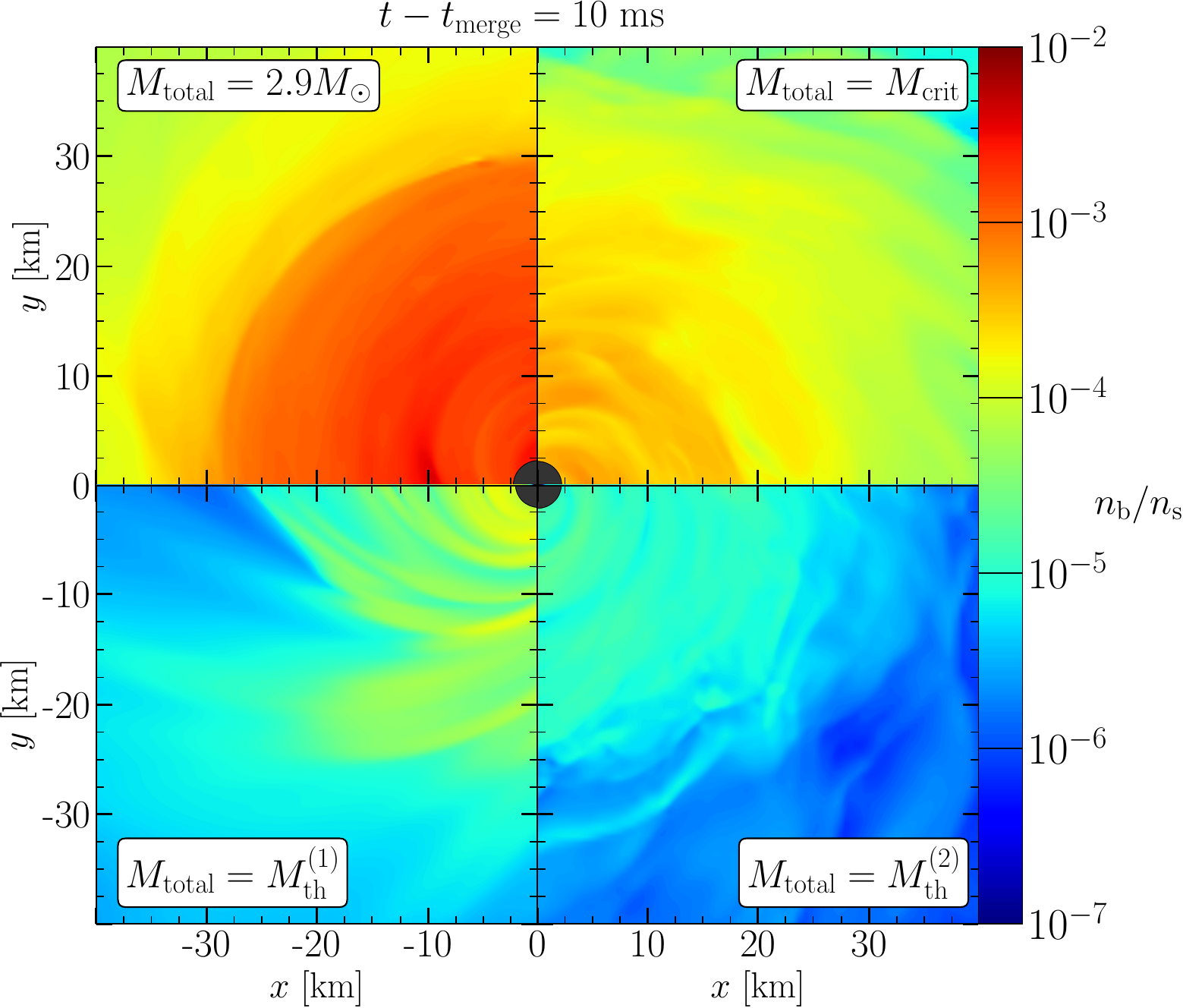}
\caption{\label{fig:K1}Left: maximum of the Kretschmann scalar as a function of the total mass and time elapsed after the merger. Right: Density in the torus ten milliseconds after the merger. Both figures from~\cite{Ecker:2024kzs}.
}
\end{figure}

We argued that a good observable for precisely identifying the threshold masses is one of the curvature invariants, the Kretschmann scalar
 $K_1 = R_{abcd}R^{abcd}$,
where $R_{abcd}$ is the Riemann tensor. We defined the critical mass
\begin{equation}
 M_{\rm crit} = {\rm min}(M): \quad \frac{dt_{\rm crit}}{dM_{\rm total}} < 0 \quad \forall M_{\rm total}>M
\end{equation}
where $t_{\rm crit}$ is the time of the formation of an apparent horizon and $M_{\rm total}$ is the total mass of the binary. As one can see from the left panel in Fig.~\ref{fig:K1}, the critical mass (purple curve) separates mergers where the hypermassive star either collapses directly to a black hole, or bounces once before eventually collapsing. 
Further, we defined and a sequence of threshold masses 
\begin{equation}
 M_{\rm th}^{(p)} = {\rm min}(M_{\rm total}): \quad \frac{d^p}{dt^p }\max (K_1) > 0 \quad \forall t>t_{\rm merge} \ ,
\end{equation}
where $p=1$, 2, 3 \ldots. Therefore the threshold masses characterize how monotonically the collapse takes place. 
The masses satisfy $M_{\rm crit} < M_{\rm th}^{(1)}< M_{\rm th}^{(2)}< \cdots$. 

As it turns out, the gravitational wave signal is rather weakly dependent on the total mass in the range of masses we study. However, the analysis suggests that the electromagnetic signal strongly depends on the mass. This is seen by studying the residual mass in the torus after the collapse (see right panel in Fig.~\ref{fig:K1}). The residual mass sharply drops for mergers above the critical mass. This drop is enhanced by the nuclear to quark matter transition~\cite{Ecker:2024kzs}, but remains even in the absence of quark matter. Since the electromagnetic signal is driven by the residual mass, this observation suggests that $M_{\rm crit}$ can be precisely measured by studying it.

\section{Conclusions}
 
I discussed how the gauge/gravity duality can be used to derive predictions for dense matter and neutron star mergers. I focused on a specific approach based on the holographic V-QCD model. With this approach, many details work remarkably well: lattice data for QCD thermodynamics at low densities can be fitted accurately, the extrapolated EOS at high densities is feasible, the nuclear matter EOS is stiff so that passing observational bounds is easy, and simultaneous description of nuclear and quark matter allows one to derive predictions for the phase transition between the phases. In addition, results for the EOS and the critical point in the phase diagram agree remarkably well with a completely different approach, namely the functional renormalization group method.

These successes of the model allowed us to construct a hybrid overall state-of-the-art EOS model for QCD by combining the predictions from V-QCD and other models, i.e., a van der Waals setup and the HS(DD2) model. The hybrid EOS was consequently used to study quark matter production and the promptness of collapse to a black hole  in neutron star mergers.

There are several ways to improve the model. A specific observation is that the model suffers from a modulated instability linked to QCD anomalies~\cite{Nakamura:2009tf}. Interestingly, unstable region extends to low densities and high temperatures in the phase diagram, potentially cloaking the critical point, and the result is universal: apart from V-QCD, the result applies to other holographic models fitted to lattice QCD data such as the Einstein-Maxwell-dilaton models~\cite{DeWolfe:2010he,Critelli:2017oub,Cai:2022omk}. Solving the nature of the inhomogeneous end point of this instability and including the phase in the EOS model will be done in future. Such an inhomogeneity may also compete with other phases in the low temperature and high density region, such as color superconducting phases (see, e.g.,~\cite{Basu:2011yg,BitaghsirFadafan:2018iqr,Preau:2025ubr}). 
 
In addition two including new phases, the description of the nuclear and (uncondensed) quark matter phases has room for improvements. These include proper implementation of quark flavors and in particular the mass of the strange quark. This also allows one to study the deviation from isospin symmetric configuration~\cite{Kovensky:2021ddl,Bartolini:2022gdf}, which can be done both for nuclear and quark matter.

\bibliographystyle{JHEP}
\bibliography{refs}

%

\end{document}